\documentclass[11]{article}

\usepackage{amsfonts}
\usepackage{amsmath}
\usepackage{enumerate}
\usepackage{epsf}
\def\graph#1#2{\vcenter{\vbox{\epsffile{graph#1.ps}}}}
                                                             
\usepackage{graphicx}

\begin{document}




\null\vskip-24pt

\hfill KL-TH 00/12



\vskip0.3truecm

\begin{center}

\vskip 3truecm

{\Large\bf

Conformal partial wave analysis of AdS amplitudes for dilaton-axion four-point functions
}\\ 

\vskip 1.5truecm


{\large\bf L. Hoffmann} \footnote{email:{\tt hoffmann@physik.uni-kl.de}}{\large\bf, L. Mesref }\footnote{email:{\tt lmesref@physik.uni-kl.de}}{\large\bf, W.  R\" uhl} \footnote{email:{\tt ruehl@physik.uni-kl.de}}

\vskip 1truecm


{\it Department of Physics, Theoretical Physics\\

University of Kaiserslautern, Postfach 3049 \\

67653 Kaiserslautern, Germany}\\

\end{center}
\vskip 1truecm

\centerline{\bf Abstract}
Operator product expansions are applied to dilaton-axion four-point functions. In the expansions of the bilocal fields $\tilde{\Phi}\tilde{\Phi}$, $\tilde{C}\tilde{C}$ and $\tilde{\Phi}\tilde{C}$, the conformal fields which are symmetric traceless tensors of rank $l$ and have dimensions $\delta=2+l$ or $8+l+\eta(l)$ and $\eta(l)=\mathcal{O}(N^{-2})$ are identified. The unidentified fields have dimension $\delta=\lambda+l+\eta(l)$ with $\lambda\geq 10$. The anomalous dimensions $\eta(l)$ are calculated at order $\mathcal{O}(N^{-2})$ for both $2^{-\frac{1}{2}}(-\tilde{\Phi}\tilde{\Phi}+\tilde{C}\tilde{C})$ and  $2^{-\frac{1}{2}}(\tilde{\Phi}\tilde{C}+\tilde{C}\tilde{\Phi})$ and are found to be the same, proving $U(1)_Y$ symmetry. The relevant coupling constants are given at order $\mathcal{O}(1)$.



\newpage

\section{Introduction}

Conformal quantum field theories are a special class of quantum field theories that has attracted a lot of interest in the last thirty years  because of its direct applicability in elementary particle and solid state physics. Solvability of these theories has often been claimed but materialized only in the case of two-dimensional theories. In more than two dimensions only the perturbative neighborhoods of free field theories have been accessible.

The situation has changed drastically after the AdS/CFT correspondence has been discovered \cite{maldacena}. The strong coupling domain of certain conformal field theories appears as holographic image of a perturbative supergravity theory which is itself derived from some string theory by compactification and separation of compactified variables.

Dilaton-axion four-point functions are simple examples to study conformal partial wave (CPW) decomposition or, what is almost the same, operator product expansion (OPE) on the holographic image of AdS field theory \cite{dobrev}. This image is a conventional conformal field theory and therefore applicability of these techniques is beyond any reasonable doubt.
In the present work we exploit these techniques in the AdS$_5$ $\rightarrow$ SYM$_4$ case to generate infinite towers of fields of SYM$_4$,to prove their non-existence in the strong t'Hooft coupling domain of SYM$_4$ or to test $U(1)_Y$ symmetry. The conformal fields (conformal blocks or quasiprimary fields) are characterized by their dimension (the anomalous part in particular), their spin ( the rank of traceless symmetric tensors) and their parity. In addition we give the coupling or fusion constants.

The holographic image of the dilaton $\tilde{\Phi}(x)$ and the axion $\tilde{C}(x)$ are chiral primary operators and have dimension $4$ and $Y$ charge $-4$. They are scalar respectively pseudo-scalar. We apply OPE to the bilocal operators
\begin{align} \label{1.1}
&\tilde{\Phi}(x_1)\tilde{\Phi}(x_3) \notag \\
&\tilde{C}(x_1)\tilde{C}(x_3) \notag \\
&\tilde{\Phi}(x_1)\tilde{C}(x_3)
\end{align}
which in the SYM$_4$ terminology are ``double-trace operators''. Each conformal block contained in them may assume an anomalous dimension. The first two bilocal fields in (\ref{1.1}) are degenerate with respect to all quantum numbers at leading order $\mathcal{O}(N^0)$ but at $\mathcal{O}(\frac{1}{N^2})$ this degeneracy is lifted by rotation of an angle
\begin{equation} \label{1.2}
\frac{\pi}{4} + \mathcal{O}(\frac{1}{N^2})
\end{equation}
so that
\begin{equation} \label{1.3}
\Psi_{\pm}(x_1, x_3) = \frac{1}{\sqrt{2}}[\pm \tilde{\Phi}(x_1)  \tilde{\Phi}(x_3)+ \tilde{C}(x_1)  \tilde{C}(x_3)]
\end{equation}
are the correct bilocal fields to start with the CPW analysis. Let us denote the last bilocal field (\ref{1.1}) after symmetrization by $\Psi_{0}(x_1, x_3)$.
Only $\Psi_{+}(x_1, x_3)$ couples to the energy-momentum tensor but $\Psi_{-,0}(x_1, x_3)$ do not. Among the towers of exceptional conformal fields of dimension
\begin{equation} \label{1.4}
\delta_l = 2 +l
\end{equation}
and even tensor rank $l \, (l\geq 2)$ which are conserved currents, only the energy-momentum tensor $l=2$ appears in $\Psi_+(x_1, x_3)$. This is different for the free-field limit of SYM$_4$, where all tensors $l$ are present and couple with (squared) coupling constants (see section 3)
\begin{equation} \label{1.5}
\gamma_l = \frac{1}{2}(2l+1)(l-1)_4, \quad l \, \text{even}.
\end{equation}
So these tensors with $l\geq4$ decouple at strong t'Hooft coupling and, according to the standard interpretation, arise from string mode excitations.

After subtraction of the energy-momentum tensor from $\Psi_{+}(x_1, x_3)$, the remaining conformal blocks in either $\Psi_{+,-,0}(x_1, x_3)$ are tensors of rank $l$, $l$ even, with dimension
\begin{equation} \label{1.6}
\delta(l,t) = 8+2t+l+\eta(l,t), \quad t \in \bf{N}_0.
\end{equation}
The exchange amplitudes of dilaton and axion also contribute only to the blocks (\ref{1.6}), obviously due to the derivative couplings.
We have analyzed the cases $t=0, l$ arbitrary, and found equal behavior for $\Psi_{-}(x_1, x_3)$ and $\Psi_{0}(x_1, x_3)$: The coupling constants are (to the order $\mathcal{O}(1)$)
\begin{equation} \label{1.7}
\epsilon(l,0) = \frac{(2l+7)(l+1)_6}{18}
\end{equation}
and the anomalous dimensions (to the order $\mathcal{O}(N^{-2})$)
\begin{equation} \label{1.8}
\eta(l,0) = -\frac{96}{(l+1)(l+6)}\frac{1}{N^2}
\end{equation}
(For $l=0$ this result was known before, \cite{dhoker1}).
The equal behavior of $\Psi_{-}(x_1, x_3)$ and $\Psi_{0}(x_1, x_3)$ in four-point functions 
\begin{equation} \label{1.9}
<\Psi_{-}(x_1, x_3)\Psi_{-}(x_2, x_4)> = <\Psi_{0}(x_1, x_3)\Psi_{0}(x_2, x_4)>
\end{equation}
follows from the $U(1)_Y$ ``bonus symmetry'' postulated in \cite{intriligator}. In fact, (\ref{1.9}) is a consequence of 
\begin{equation} \label{1.10}
<\mathcal{O}_\tau(x_1)\mathcal{O}_\tau(x_2)\mathcal{O}_\tau(x_3)\mathcal{O}_\tau(x_4)> = 0
\end{equation}
with
\begin{equation} \label{1.11}
\mathcal{O}_\tau(x)= \tilde{\Phi}(x)+i \tilde{C}(x).
\end{equation}
So our results strongly support this suggested symmetry of four-point functions.
There is another observation in \cite{arutyunov}. The tower of tensor fields (\ref{1.6}) with $t=-2$ exists also in SYM$_4$ and can be derived (\cite{arutyunov}) from two lowest $R$-symmetry weight operators
\begin{equation} \label{1.12}
Tr\{\phi^i \phi^j-\frac{1}{6} \delta_{ij} \phi^k \phi^k\}(x)
\end{equation}
($SO(6)$ traceless symmetric tensors with $SU(4)$ weight $[0,2,0]$) by operator product expansion and projection on the  $R$-symmetry representation $[0,0,0]$. At $l=0$ the anomalous dimension has been calculated in \cite{arutyunov} to be
\begin{equation} \label{1.13}
\eta(0,-2) = -\frac{16}{N^2} = \eta(0,0)
\end{equation}
which coincides with $\eta(0,0)$ in (\ref{1.8}). This is explained by the fact that the field $(l,t)=(0,0)$ in (\ref{1.6}) is a supercharge-descendant of the field  $(l,t)=(0,-2)$, whereby the $Y$ charge of the latter, namely zero, is lowered to $Y=-8$. It is $Y$ charge conservation $(U(1)_Y)$ which forbids production of the towers $(l,-2)$ and $(l,-1)$ from pairs of dilatons and axions. Moreover, (\ref{1.13}) should be generalized to
\begin{equation} \label{1.14}
\eta(l, -2) = \eta(l, 0)
\end{equation}
which will be verified in a forthcoming paper.

Remarkably, the coupling constants (\ref{1.7}) appear also in the free-field limit at order $\mathcal{O}(1)$. The explanation is that they are derived from disconnected graphs containing only propagators of chiral primary operators and these satisfy nonrenormalization theorems. At order $\mathcal{O}(N^{-2})$ both sequences of coupling constants should be different.
Of course the odd tensor ranks contained in 
\begin{equation} \label{1.15}
\tilde{\Phi}(x_1) \tilde{C}(x_3)
\end{equation}
are independent. We will treat them as well.

We do not evaluate any AdS graphs but rely on the work \cite{dhoker1, freedman, dhoker2} . The four-point function of the local field
\begin{equation} \label{1.16}
:F_{\mu\nu}(x)F_{\mu\nu}(x):
\end{equation}
is taken from \cite{herzog}.
The CPW analysis of this work is based on the formalism developed in \cite{lang1}. We have to adjust it to spacetime dimension four and make it more explicit, in particular for the exceptional series of field representations in Section 2. In Section 3 we submit the four-point function of the field (\ref{1.16}) in the free-field limit to a CPW analysis. In Section 4 we express the $\mathcal{O}(\frac{1}{N^2})$ contributions to the four-point functions by two-variable hypergeometric functions and extract the conformal block of the energy-momentum tensor. From the remaining  towers of conformal blocks (\ref{1.6}), we treat only the case $t=0$ in Section 5, where we derive the anomalous dimensions (\ref{1.8}).

All our arguments in this paper are analytic and therefore it was necessary to enlarge the known analytic formalism to a certain extent. We present the corresponding material in three appendices. In Appendix A we give explicit formulas for the contributions of the derivative fields to the conformal blocks of the exceptional repesentations (up to the third derivative). For the other representations we were unable to simplify the general expressions (2.8), (2.10). In Appendix B we derive analytic continuations (''Kummer formulae'') of the two-variable hypergeometric functions from the symmetry of the graphs. This could also be of mathematical interest. Finally we give the proof of the formula of the double-trace fields (5.42) (or (\ref{1.8})) in Appendix C.  

\section{Conformal partial wave analysis}
\setcounter{equation}{0}
Conformal partial wave analysis applied to a conformally covariant $n$-point function is a concept derived from harmonic analysis on the conformal group.
However, for a conformal quantum field theory with fields transforming as "elementary" representations, which are characterized by a dimension $\delta$ and a spin transformation behavior $(j_1, j_2), 2j_i \in \bf{N}_0$, the conformal partial waves can be replaced by exchange Green functions for conformal fields. An $n$-point function can be decomposed in a channel
\begin{equation} \label{2.1}
(1,2,...,m) \longrightarrow (m+1, m+2,...,n)
\end{equation}
into the covergent sum
\begin{multline} \label{2.2}
<\Phi_1(x_1) \Phi_2(x_2)...\Phi_n(x_n)>  \\= \sum_\Psi \gamma_{\{1,2,...,m\}}^\Psi(A) \gamma_{\{m+1,m+2,...,n\}}^\Psi(B) \graph{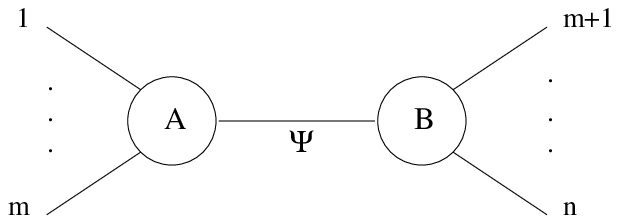}{1}
\end{multline}
where $\Psi$ runs over all conformal fields that couple to the fields $(1,2,...,m)$ and $(m+1, m+2,...,n)$ simultaneously and $\gamma^{\Psi}_{1,2,...,m}(A)$ is a "coupling constant" for the vertex function $A$, $\gamma^{\Psi}_{1,2,...,m}(B)$ correspondingly for $B$.
In the $\Psi$-exchange amplitudes the shadow terms must be skipped. The derivative fields of $\Psi$ are automatically included in the exchange amplitude, which implies that actually a "conformal block" is exchanged.
In the case of a four-point function of four scalar fields with pairwise equal dimensions
\begin{equation} \label{2.3}
\delta_1 = \delta_2, \quad \delta_3 = \delta_4
\end{equation}
the exchange fields are symmetric traceless tensor fields with dimension
\begin{equation} \label{2.4}
\delta = \lambda + l,
\end{equation}
where $l$ is the rank of the field and is, in terms of $(j_1, j_2)$, expressed by
\begin{equation} \label{2.5}
j_1 = j_2 = \frac{1}{2}l.
\end{equation} 
Unitarity requires \cite{mack}
\begin{equation} \label{2.6}
\lambda \geq \begin{cases}
  2,\quad l \neq 0, \\
  1,\quad l = 0
\end{cases}
\end{equation}
for spacetime dimension $d = 4$.
The case
\begin{equation} \label{2.7}
\lambda = 2,\quad l \neq 0
\end{equation}
is "exceptional" and belongs to conserved tensor currents.
The exchange amplitude for arbitrary tensor rank $l$ was derived in \cite{lang1} and is, without shadow term and in the channel $(1,3) \rightarrow (2,4)$, given by
\begin{equation} \label{2.8}
\gamma_{1 3}^{\Psi}\gamma_{2 4}^{\Psi} (x_{1 2}^2)^{-\delta_1} (x_{3 4}^2)^{-\delta_3} \,\sum_{M \in \{l, l-2, l-4,...0\}} 2^{-l} \frac{G_M^{(l)}}{G_l^{(l)}} \, u^{\frac{1}{2}(\delta-\delta_1-\delta_3-M)}  \sum_{n, m=0}^\infty \, \frac{u^n (1-v)^m}{n! m!} \, A_{nm}^{(M)}(\delta_1, \delta_3, \lambda, l)
\end{equation}
with
\begin{equation}
u=\frac{x_{13}^2x_{24}^2}{x_{12}^2x_{34}^2}, \; v=\frac{x_{14}^2x_{23}^2}{x_{12}^2x_{34}^2}, \; x_{ij} = x_i -x_j. \notag
\end{equation}  
Here $G_M^{(l)}$ are the coefficients of the Gegenbauer polynomial $C^{\frac{1}{2}d-1}_l(t)$
\begin{equation} \label{2.9}
 C^{\frac{1}{2}d-1}_l(t) = \sum_M \, G_M^{(l)} t^M; \quad G_M^{(l)} = 0 \quad\text{if}\quad l - M \quad \text{is odd}
\end{equation}
and 
\begin{align} \label{2.10}
A_{nm}^{(M)} =& \pi^{\frac{d}{2}} (-1)^{n+M} \sum_{r, s, t} (-1)^{r+s+t} \binom{n}{s} \binom{n+s}{t} \binom{n-s}{r} \frac{M! (m+n+s-t)!}{(M-t-r)! (m+n+s-M)!} \notag \\ &\times \Gamma(A_3 +\frac{1}{2}M+n-r-t) \Gamma(A_3-\frac{1}{2}M+m+n+s)\Gamma(A_1-\frac{1}{2}M+m+n) \notag \\
&\times\Gamma(A_2 +A_4+M-t-r-1) \Gamma(A_4-A_3-n) 
\{\Gamma(A_2+A_4-1) \notag \\ &\times \Gamma(A_1+A_3+m+n+s-t) \Gamma(A_2+\frac{1}{2}M) \Gamma(A_3+\frac{1}{2}M) \Gamma(A_4+\frac{1}{2}M-t-r)\}^{-1}
\end{align}
where furthermore
\begin{equation} \label{2.11}
A_1 = \frac{d}{2} - A_4 = \frac{1}{2}(\delta + \delta_1 - \delta_3), \quad
A_3 = \frac{d}{2} - A_2 = \frac{1}{2}(\delta - \delta_1 + \delta_3) 
\end{equation}
so that
\begin{equation} \label{2.12}
A_1 + A_3 = \delta, \quad A_2 + A_4 = d - \delta.
\end{equation}

These formulae are extremely complicated, which entails that analytic proofs are often very difficult. We are interested, in the present work, only in special cases
\begin{equation} \label{2.13}
\delta_1 = \delta_3 = 4, \quad d = 4
\end{equation}
and integer values of $\lambda$. In this case the formulae reduce in size.
Let us extract a factor first
\begin{equation} \label{2.14}
A_{nm}^{(M)} = \gamma_{13}^{\Psi} \gamma_{24}^{\Psi}\; \pi^{\frac{d}{2}} 2^{-l} \frac{\Gamma(d-\lambda-1) \Gamma(\frac{d}{2}-\lambda-l)}{\Gamma(d-\lambda-l-1) \Gamma(\frac{d}{2}-\frac{\lambda}{2})^2}\; \tilde{A}_{nm}^{(M)}.
\end{equation}
Then $\tilde{A}_{nm}^{(M)}$ can be given as \footnote{$l-M$ is even}
\begin{align} \label{2.15}
\tilde{A}_{nm}^{(M)} =& (-1)^{n+l} \sum_{r, s, t} (-1)^{r+s+t} \binom{n}{s} \binom{n+s}{t} \binom{n-s}{r} \frac{M! (m+n+s-t)!}{(M-t-r)! (m+n+s-M)!} \notag \\
&\times\Gamma(\frac{1}{2}(\lambda+l+M)+n-r-t) \Gamma(\frac{1}{2}(\lambda+l-M)+m+n)  \notag \\ \times &\Gamma(\frac{1}{2}(\lambda+l-M)+m+n+s) \{\Gamma(\frac{1}{2}(\lambda+l+M)) \Gamma(\lambda+l+m+n+s-t)\}^{-1} \,  F_A\cdot F_B \cdot F_C
\end{align}
where $F_A, F_B, F_C$ are three singular factors. Each of the factors is equal one if (see below)
\begin{equation} \label{2.16}
M = l, n = 0 \quad (\text{implying}\quad r = s = t = 0)
\end{equation}
and 
\begin{equation} \label{2.17}
\tilde{A}_{0m}^{(l)} = \frac{(-1)^l m!}{(m-l)!} \frac{\Gamma(\frac{1}{2}\lambda+m)^2}{\Gamma(\lambda+l+m)}
\end{equation}
is constant in $d$ and can trivially be continued to the values (\ref{2.13}) and $\lambda \in \bf{N}$.
The singular factor $F_A$ is
\begin{align} 
F_A =& \frac{\Gamma(\frac{d}{2}-\lambda-l-n)}{\Gamma(\frac{d}{2}-\lambda-l)} \label{2.18} \\
&\underset{(d \rightarrow 4)}{\longrightarrow} (-1)^n \frac{\Gamma(\lambda+l-1)}{\Gamma(\lambda+l+n-1)} \label{2.19}
\end{align}
which is finite except for the exceptional case (\ref{2.6})
\begin{equation} \label{2.20}
l = 0, \; \lambda = 1
\end{equation}
which will not occur.
The singular factors $F_B, F_C$ are
\begin{align}
F_B =& \frac{\Gamma(d-\lambda-(l-M)-t-r-1)}{\Gamma(d-\lambda-1)} \label{2.21} \\
F_C =& \frac{\Gamma(\frac{d}{2}-\frac{\lambda}{2})^2}{\Gamma(\frac{d}{2}-\frac{1}{2}(\lambda+l-M))\Gamma(\frac{d}{2}-\frac{1}{2}(\lambda+l-M)-t-r)} \label{2.22}
\end{align}
and depend on $\lambda$ sensitively. We distinguish the two cases
\begin{align}
&(\alpha) \quad \lambda = 2, \label{2.23} \\
&(\beta) \quad \lambda \geq 3. \label{2.24}
\end{align}
In the case $(\alpha)$ $F_B$ has a first order pole in $d$ if 
\begin{equation} \label{2.25}
(l-M)+t+r \geq 1 \quad (\in \bf{N})
\end{equation}
whereas $F_C$ has
\begin{equation}
\begin{cases}
\text{if}\quad l-M > 0: \, \text{a second order zero} \notag \\
\text{if}\quad l-M = 0, t+r \geq 1: \text{a first order zero} \notag
\end{cases}
\end{equation}
in $d$ at $d=4$. Thus for case $(\alpha)$ only
\begin{equation} \label{2.26}
 l-M = 0,\; t+r \geq 1
\end{equation}
gives a finite non-vanishing contribution and
\begin{equation} \label{2.27}
\underset{d \rightarrow 4}{\text{lim}} F_B \cdot F_C = \frac{1}{2}(1+ \delta_{t+r,0}).
\end{equation}
The second term comes from $l-M=t+r=0$, where no pole and no zero is present.
Thus we conclude, that in the case of the exceptional representations (\ref{2.23}) only the leading power of the Gegenbauer polynomial survives.
In the case $(\beta)$ both $F_B$ and $F_C$ have finite limits at $d=4$
\begin{equation} \label{2.28}
\underset{d \rightarrow 4}{\text{lim}} F_B \cdot F_C = \frac{(\frac{1}{2}(\lambda-2))_{\frac{1}{2}(l-M)} (\frac{1}{2}(\lambda-2))_{\frac{1}{2}(l-M)+r+t}}{(\lambda-2)_{l-M+r+t}}.
\end{equation}

Now we define
\begin{equation} \label{2.29} 
\sum_{M \in \{l,l-2,l-4,...\}} \frac{G_M^{(l)}}{G_l^{(l)}} u^{\frac{1}{2}(l-M)} \sum_{n,m=0}^{\infty} \tilde{A}_{nm}^{(M)} \frac{u^n (1-v)^m}{n! m!}  = \sum_{n=0}^{\infty} \frac{u^n}{n!} G_n(\lambda, l; v). 
\end{equation}
Then we get from (\ref{2.17})
\begin{equation} \label{2.30}
G_0(\lambda, l; v) = \frac{\Gamma(\frac{1}{2}\lambda+l)^2}{\Gamma(\lambda+2l)}(v-1)^l \;_2F_1(\frac{1}{2}\lambda+l, \frac{1}{2}\lambda+l; \lambda+2l; 1-v).
\end{equation}
This expression is valid in the case of general $(\lambda,\, l)$. But the functions $G_n, \, n \geq 1$ are so clumsy, that we prefer to give them only in the exceptional case $\lambda = 2$.
Then we get
\begin{equation} \label{2.31}
 G_n(2, l; v) = \sum_{m=0}^{\infty} \tilde{A}_{nm}^{(l)}(\lambda=2) \frac{(1-v)^m}{m!}
\end{equation}
with
\begin{align} 
\tilde{A}_{nm}^{(l)}(\lambda=2) = \frac{(-1)^l}{2(l+1)_n} \sum_{t,s} (-1)^{t+s} \binom{n}{s} \binom{n+s}{t} &\frac{(m+n)! (m+n+s)! (m+n+s-t)!}{(m+n+s-l)! (m+n+s+l+1-t)!} \notag \\
&\times \{ \frac{n!}{s!} (l+1-t)_s +(l+1)_n \delta_{t,0}\}. \label{2.32}
\end{align}
Performing the $t$-summation leads to a group of terms for each $s$. The lengthy results for the cases $n \in \{1, 2, 3\}$ are given explicitly in Appendix A.

\section{The free field limit of the dilaton four-point function and its conformal partial wave decomposition}
\setcounter{equation}{0}
AdS-CFT correspondence maps type $IIB$ supergravity theory on Super-Yang-Mills theory with $\mathcal{N}=4$ supercharges and gauge group $SU(N)$. In this conformal gauge theory the local field
\begin{equation} \label{3.1}
F^2(x) = \,:\sum_A F_{\mu\nu}^A(x)  F_{\mu\nu}^A(x):
\end{equation}
appears as image of the dilaton field. The normal product (\ref{3.1}) is defined, as usual, perturbatively. The index "$A$" runs over an orthogonal basis in $(N^2-1)$-dimensional Lie algebra space $su(N)$, so that, $F_{\mu\nu}^A(x)$ are real fields. In the strong coupling domain where the AdS-CFT image of the dilaton must be looked for, the normal product (\ref{3.1}) should be defined nonperturbatively. Then the correspondence still holds since the operator (\ref{3.1}) is "chiral primary" and its anomalous dimension vanishes for all coupling constants $g^2_{YM}N$.
Some of the calculations of this section have been done by Herzog \cite{herzog} (see also \cite{arutyunov}, \cite{dolan}). 
We consider
\begin{equation} \label{3.2}
M_4(x_1,x_2,x_3,x_4) = <F^2(x_1)F^2(x_2)F^2(x_3)F^2(x_4)>
\end{equation}
and apply a conformal partial wave analysis in the channel
\begin{equation} \label{3.3}
(1, 3) \longleftrightarrow (2, 4).
\end{equation}

In the free field limit we can calculate
\begin{align} 
W_{12}(x_1,x_2)= &\sum_{A, B} <F_{\mu\nu}^A(x_1)  F_{\lambda\sigma}^B(x_2)>  <F_{\lambda\sigma}^B(x_2)  F_{\mu\nu}^A(x_1)> \label{3.4} \\
&= \frac{24}{x^8_{1 2}} \gamma^2 g_{YM}^4 (N^2-1) \label{3.5}
\end{align}
with $\gamma^\frac{1}{2}$ being a numerical field normalization factor and
\begin{align} 
W_{1324} =& \sum_{A,B,C,D} <F_{\mu\nu}^A(x_1)  F_{\rho\tau}^C(x_3)>  <F_{\rho\tau}^C(x_3)  F_{\lambda\sigma}^B(x_2)> \notag \\
&\times <F_{\lambda\sigma}^B(x_2) F_{\xi\eta}^D(x_4)>  <F_{\xi\eta}^D(x_4) F_{\mu\nu}^A(x_1)> \\
&= \frac{32}{x_{12}^8 x_{34}^8} \gamma^4 g_{YM}^8 (N^2-1) [\frac{(u+v-1)^2-uv}{(uv)^3}] \label{3.7}
\end{align}
with $u, v$ as in (\ref{2.8}). Then
\begin{equation} \label{3.8}
M_4 = 4 (W_{12} W_{34} + W_{13} W_{24}+ W_{14} W_{23}) + 16 (W_{1234} + W_{1324} + W_{1342}).
\end{equation}
The term $4 W_{13} W_{24}$ gives the unit operator contribution to the operator product expansion of $M_4$ in the $(1, 3) \longleftrightarrow (2, 4)$ channel. The rest 
\begin{equation} \label{3.9}
M_4 - 4  W_{13} W_{24}
\end{equation}
is normalized in such a fashion that a comparison with the AdS result (see section 4) is easily done. Then we get
\begin{align}
(x_{12}^2 x_{34}^2)^{-4} &\{ 1 + v^{-4} + \frac{2}{9 (N^2-1)} [ (uv)^{-3}((u+v-1)^2 - uv) \notag \\
&+ u^{-3}((u+1-v)^2-u) + v^{-3} ((v+1-u)^2 - v)]\}. \label{3.10}
\end{align}
We start the conformal partial wave analysis with a tower of exceptional representation fields ($\lambda = 2, l \geq 2$ \, even) contained in the $\mathcal{O}(N^{-2})$ terms in (\ref{3.10}). 
They have dimensions
\begin{equation} \label{3.11}
\delta =  2 + l
\end{equation}
which, as we know, belong to conserved tensor currents. Then we analyze a sequence of towers labelled by $t \in \bf{N}_0$ (``twist'') of tensors of rank $l$ ($l$ even, $\geq 0$) and dimension
\begin{equation} \label{3.12}
\delta = 8 +2t +l.
\end{equation}
For $t=l=0$ we get in particular the field
\begin{equation} \label{3.13}
:(F^2(x))^2:
\end{equation}
as an image of the two-dilaton field.
The square bracket in (\ref{3.10}) can be expanded into
\begin{equation} \label{3.14}
[...] = \sum_{n = -2}^3 u^{-n} S_n(v)
\end{equation}
with
\begin{align}
S_3(v) &= (1+v^{-3})(1-v)^2, \label{3.15} \\
S_2(v) &= -2v^{-3} + v^{-2} + 1 -2v, \label{3.16} \\
S_1(v) &= 1 +v^{-3} \label{3.17}
\end{align}
etc.
From (\ref{2.8}) we obtain with (\ref{3.11})
\begin{equation} \label{3.18}
u^{\frac{1}{2}(\delta-\delta_1-\delta_3-l)} = u^{-3}
\end{equation}
and from (\ref{2.29}) and (\ref{2.30})
\begin{align}
S_3 &=\sum_{l \in 2\bf{N}} \gamma_l G_0(2,l;v) \notag \\
    &=\sum_{l \in 2\bf{N}} \gamma_l \frac{(l!)^2}{(2l+1)!} (v-1)^l \, _2F_1(l+1,l+1; 2l+2;1-v) \notag \\
    &=(1-v)^2 \sum_{m=0}^\infty \frac{(1-v)^m}{m!} [(3)_m +\delta_{m,0}]. \label{3.19}
\end{align}
The solution of (\ref{3.19}) is
\begin{equation} \label{3.20}
\gamma_l = \frac{1}{2} (2l+1) (l-1)_4, \quad l \in 2\bf{N}.
\end{equation}
All $\gamma_l$ are positive as they should be as squares (!) of coupling constants. The case $l=2$ belongs to the energy momentum tensor which will turn out to be the only exceptional representation field in the strong coupling limit (see section 4).
With the $\gamma_l$ of (\ref{3.20}), we find
\begin{align}
S_2(v) &=  \sum_{l \in 2 \bf{N}} \gamma_l G_1(2, l ;v), \label{3.21} \\
S_1(v) &=  \frac{1}{2} \sum_{l \in 2 \bf{N}} \gamma_l G_2(2, l ;v) \label{3.22}
\end{align}
so that no conformal blocks with $\lambda=4$ or $\lambda=6$ arise.
At $\lambda=8$ we derive the tower of $(F^2)^2$ tensors at leading order in $N^2$. Then we have to solve for the coupling constants $\{\epsilon_l\}$ in
\begin{gather}
1 + v^{-4} = \sum_l \epsilon_l G_0(8, l; v) \notag \\
= \sum_l \epsilon_l \frac{((l+3)!)^2}{(2l+7)!} (v-1)^l \, _2F_1(l+4, l+4; 2l+8; 1-v). \label{3.23}
\end{gather}
Setting
\begin{equation} \label{3.24}
z = 1-v
\end{equation}
and differentiating (\ref{3.23}) $m$ times at $z=0$ gives
\begin{equation} \label{3.25}
\frac{(2m+7)!}{3! m! (m+3)!} (1+\delta_{m,0}) = \sum_l \epsilon_l \binom{2m+7}{m-l}.
\end{equation}
Though the $\{\epsilon_l\}$ can easily be calculated from (\ref{3.25}), it turns out to be difficult to guess the general solution while looking only at the numbers for the $\{\epsilon_l\}$.

This issue can be simplified considerably by introducing $\lambda$ as a parameter. Namely, consider the operator product expansion of a pair of scalar fields of dimension $\frac{1}{2}\lambda$ into symmetric traceless tensor fields of rank $l$ and dimension
\begin{equation} \label{3.26}
\delta = \lambda +l.
\end{equation}
Then, instead of (\ref{3.23}), we have
\begin{equation} \label{3.27}
1+v^{-\frac{\lambda}{2}} = \sum_l \epsilon_l^{(\frac{\lambda}{2}-1)} G_0(\lambda, l;v)
\end{equation}
or by (\ref{3.25}) with $k=\frac{1}{2}\lambda-1$
\begin{equation} \label{3.28}
\frac{(2m+2k+1)!}{k!m!(m+k)!} (1+\delta_{m,0}) = \sum_l \epsilon_l^{(k)} \binom{2m+2k+1}{m-l}.
\end{equation}
The two-parameter function $\epsilon_l^{(k)}$ can be guessed to be
\begin{equation} \label{3.29}
\epsilon_l^{(k)} =
\begin{cases} 
2 \frac{(2l+2k+1)(l+1)_{2k}}{(k!)^2}, \quad l \in 2\bf{N}_0  \\
0, \quad l \in 2\bf{N}_0+1
\end{cases}
\end{equation}
from which we deduce the solution of (\ref{3.23}) as
\begin{equation} \label{3.30}
\epsilon_l = \epsilon_l^{(3)} = \frac{1}{18} (2l+7)(l+1)_6, \quad l \in 2\bf{N}_0.
\end{equation}
All $\epsilon_l$ are integral multiples of $\epsilon_0$
\begin{equation} \label{3.31}
\epsilon_0=280.
\end{equation}

\section{AdS four-point functions, their graphs and the graviton exchange}
\setcounter{equation}{0}
The dilaton field $\Phi$ and axion field $C$ are holographically mapped on scalar conformal fields $\tilde{\Phi}$ and $\tilde{C}$ of dimension four and parity $\pm 1$, respectively, in SYM$_4$.
We are interested in the four-point functions
\begin{align}
A &= <\tilde{\Phi}(x_1)\tilde{\Phi}(x_3)\tilde{\Phi}(x_2)\tilde{\Phi}(x_4)>, \notag \\
B &= <\tilde{\Phi}(x_1)\tilde{\Phi}(x_3)\tilde{C}(x_2)\tilde{C}(x_4)>, \notag \\
C &= <\tilde{C}(x_1)\tilde{C}(x_3)\tilde{\Phi}(x_2)\tilde{\Phi}(x_4)>, \notag \\
D &= <\tilde{C}(x_1)\tilde{C}(x_3)\tilde{C}(x_2)\tilde{C}(x_4)> \label{4.1}
\end{align}
and 
\begin{equation} \label{4.2}
E =  <\tilde{\Phi}(x_1)\tilde{C}(x_3)\tilde{\Phi}(x_2)\tilde{C}(x_4)>.
\end{equation}
$B, C$ and $E$ are analytic continuations of each other so that only $A, B, D$ are essentially different.

The fields $\tilde{\Phi}$ and $\tilde{C}$ are normalized as 
\begin{equation} \label{4.3}
<\tilde{\Phi}(x)\tilde{\Phi}(0)>\, = \,<\tilde{C}(x)\tilde{C}(0)>\, = \,(x^2)^{-4}.
\end{equation}
The amplitudes (\ref{4.1}), (\ref{4.2}) can be expanded in powers of $N^{-2}$, e.g.
\begin{equation} \label{4.4}
A = A_0 + \frac{1}{N^2}A_1+  \frac{1}{N^4}A_2+ ... .
\end{equation}
The calculation done here includes $\mathcal{O}(N^{-2})$.
For $A_0 = D_0$, we have the contributions of the graphs of Fig.1
§§§§§§§§§§§§§§§§§§§§§§§§§§§§§§§§§§§§§§§§§§§§§§§§§§§§§§§§§§§§§§§§§§§§§
\begin{figure}[htb]
\begin{centering}
\includegraphics[scale=0.45]{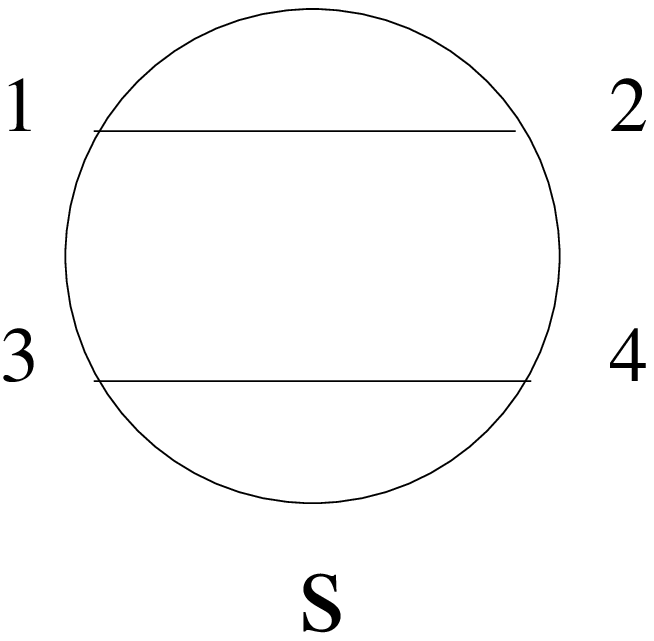}  \qquad  \includegraphics[scale=0.45]{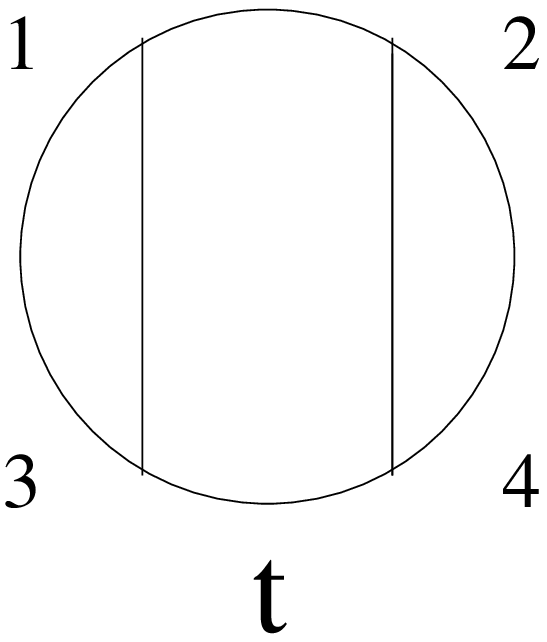}  \qquad  \includegraphics[scale=0.45]{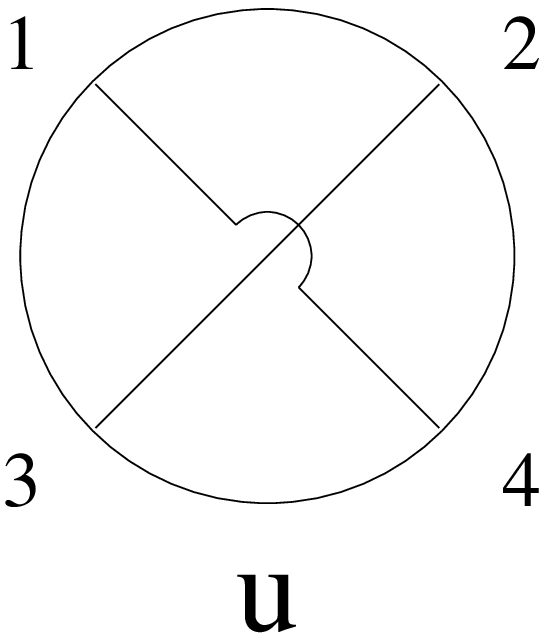}
\caption{Graphs $s, t, u$ for $A_0$ and $D_0$. For $B_0$ and $C_0$ only the graph $t$ contributes}
\end{centering}
\end{figure}
§§§§§§§§§§§§§§§§§§§§§§§§§§§§§§§§§§§§§§§§§§§§§§§§§§§§§§§§§§§§§§§§§§§§§

Since we are interested in an operator product expansion in the channel
\begin{equation} 
(1, 3) \longleftrightarrow (2, 4) \notag
\end{equation}
we can skip the (trivial) contribution of the unit operator  which comes from graph ``$t$'' in Fig.1. Then there remains
\begin{align}
A_0 = D_0 &: (x_{12}^2x_{34}^2)^{-4} (1+v^{-4}), \label{4.5}\\
B_0 = C_0 &: 0, \label{4.6}\\
E_0 &: (x_{12}^2x_{34}^2)^{-4}. \label{4.7}
\end{align}
§§§§§§§§§§§§§§§§§§§§§§§§§§§§§§§§§§§§§§§§§§§§§§§§§§§§§§§§§§§§§§§§§§§
\begin{figure}[htb]
\begin{centering}
\includegraphics[scale=0.45]{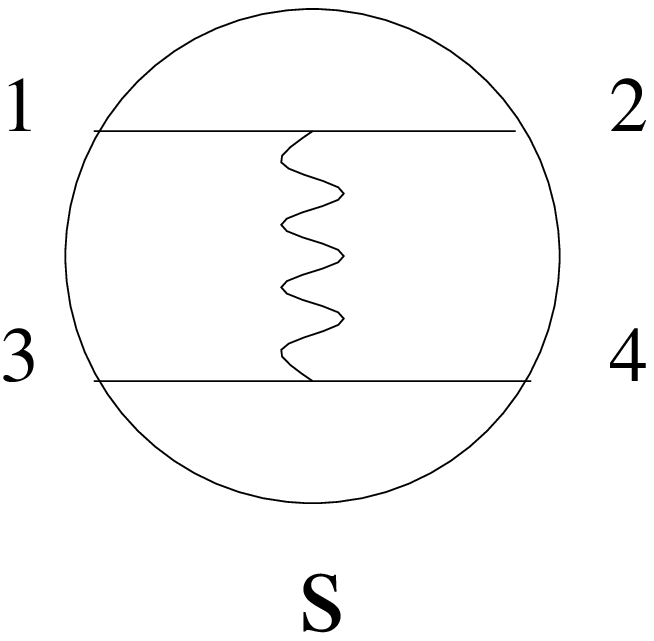} \qquad \includegraphics[scale=0.45]{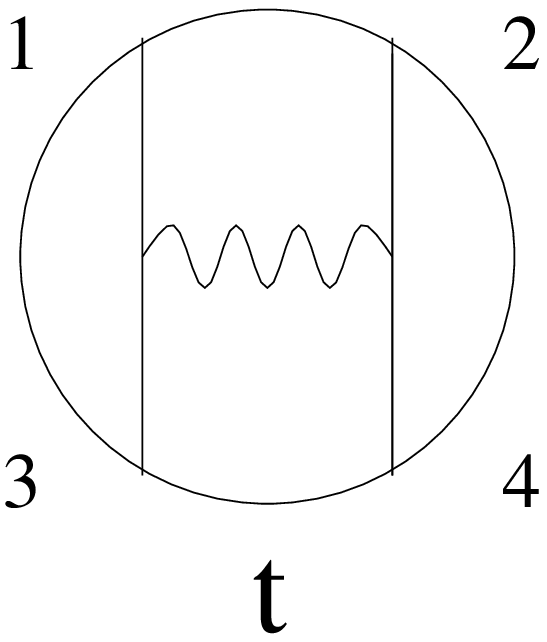} \qquad \includegraphics[scale=0.45]{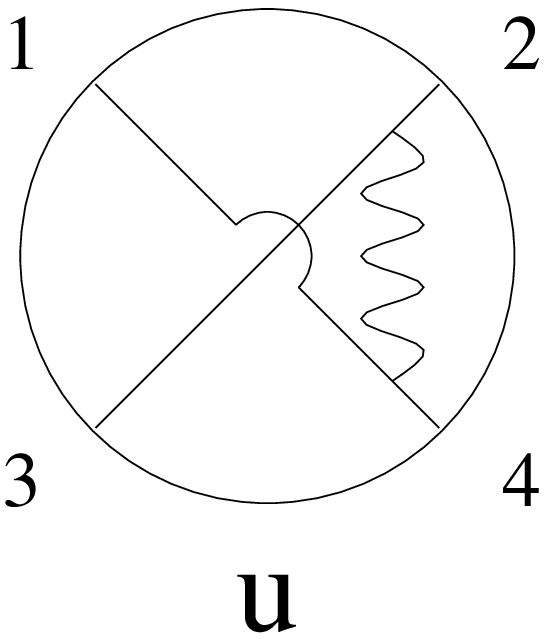}
\caption{Graviton exchange graphs $s, t, u$ for $A_1$ and $D_1$ and $t$ for $B_1$ and $C_1$  }
\end{centering}
\end{figure}
§§§§§§§§§§§§§§§§§§§§§§§§§§§§§§§§§§§§§§§§§§§§§§§§§§§§§§§§§§§§§§§§§§§§§§

At the next order $\mathcal{O}(N^{-2})$ we have the graviton exchange graphs of Fig.2 with
\begin{equation} \label{4.8}
A_1 = \frac{\pi^6}{72}[I_{grav}^{(s)}+I_{grav}^{(t)}+I_{grav}^{(u)}].
\end{equation}

In addition $D_1$ gets contributions from the dilaton exchange graphs of Fig.3, which have been shown to cancel each other \cite{freedman}. The proof of this cancellation is based on an argument by Liu and Tseytlin \cite{liu}. In the case of $B_1$, we have the axion and graviton exchange and the contact graph of Fig.4.
§§§§§§§§§§§§§§§§§§§§§§§§§§§§§§§§§§§§§§§§§§§§§§§§§§§§§§§§§§§§§§§§§§
\begin{figure}[htb]
\begin{centering}
\includegraphics[scale=0.45]{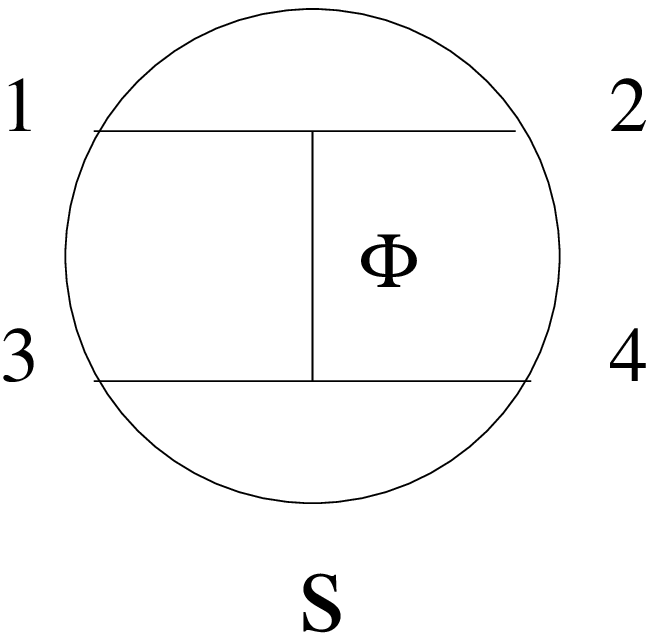} \qquad \includegraphics[scale=0.45]{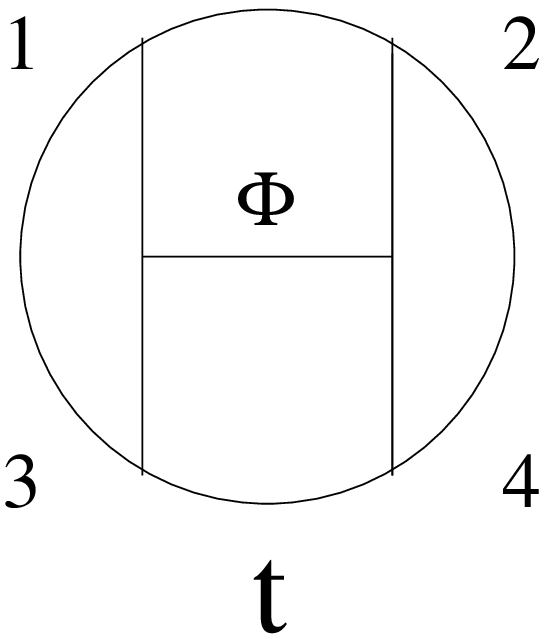}\qquad \includegraphics[scale=0.45]{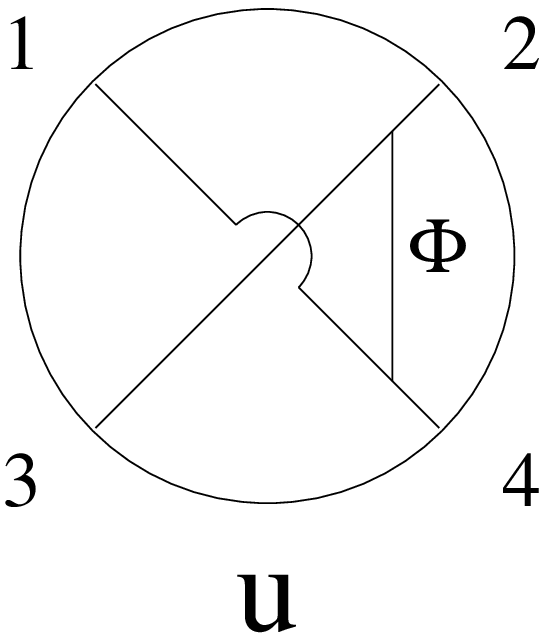}
\caption{Dilaton exchange graphs for $D_1$}
\end{centering}
\end{figure}
§§§§§§§§§§§§§§§§§§§§§§§§§§§§§§§§§§§§§§§§§§§§§§§§§§§§§§§§§§§§§§§§§§

§§§§§§§§§§§§§§§§§§§§§§§§§§§§§§§§§§§§§§§§§§§§§§§§§§§§§§§§§§§§§§§§§§§§
\begin{figure}[htb]
\begin{centering}
\includegraphics[scale=0.45]{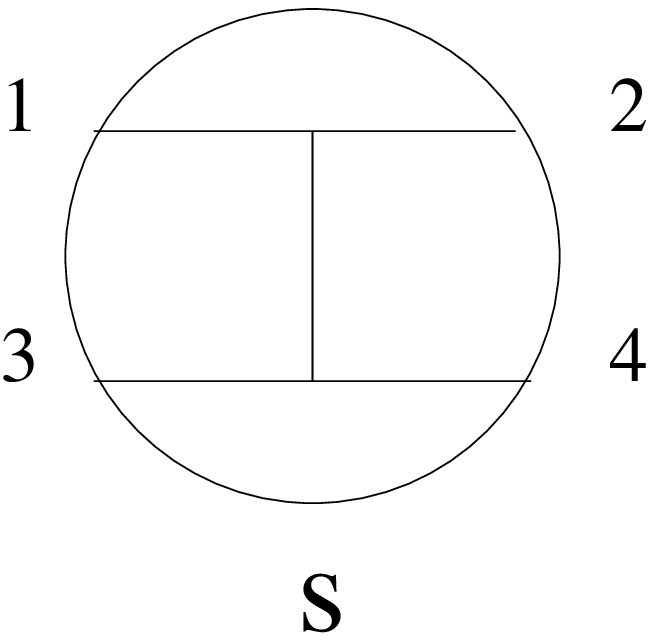}\qquad \includegraphics[scale=0.45]{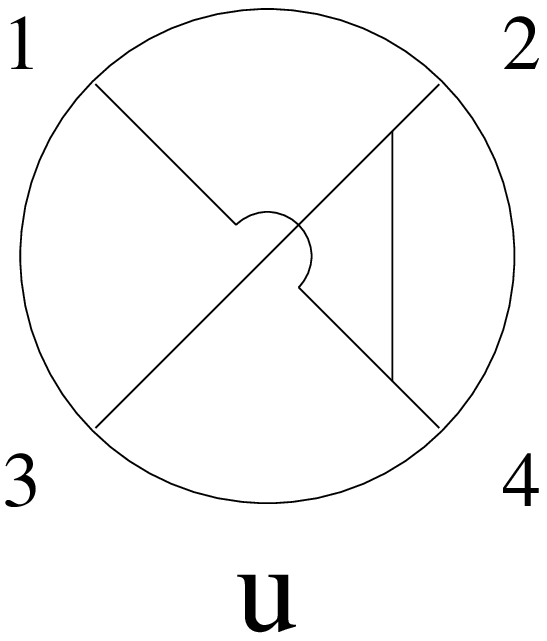} \qquad \includegraphics[scale=0.45]{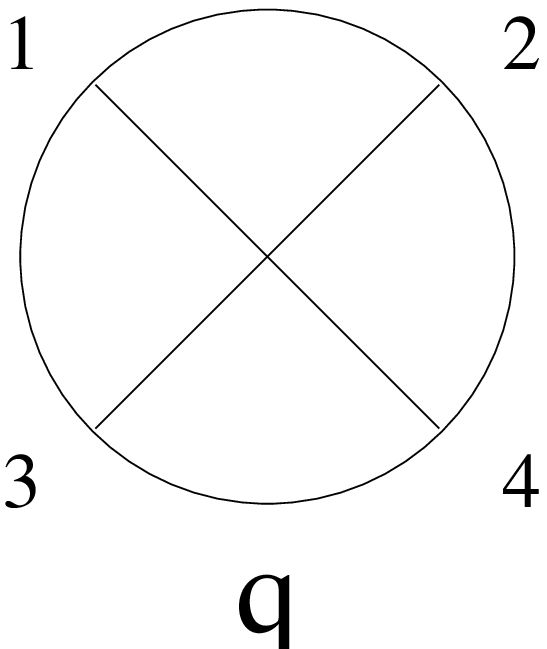}  \qquad \includegraphics[scale=0.45]{p4fig4.eps}
\caption{The graphs for $B_1=C_1$}
\end{centering}
\end{figure}

§§§§§§§§§§§§§§§§§§§§§§§§§§§§§§§§§§§§§§§§§§§§§§§§§§§§§§§§§§§§§§§§
The contribution is 
\begin{equation} \label{4.9}
B_1 = C_1 = \frac{\pi^6}{72}[I_{grav}^{(t)} + I^{(s+u+q)}].
\end{equation}
All the integrals $I$ have been evaluated \cite{dhoker2}. It was possible to express them in terms of AdS four-star functions
\begin{equation} \label{4.10}
D_{\Delta_1 \Delta_2 \Delta_3 \Delta_4}(x_1,x_2,x_3,x_4) = \int_{z_0>0} \frac{d^5z}{z_0^5} \prod_{i=1}^4 \frac{z_0^{\Delta_i}}{(z_0^2+(\vec{z}-\vec{x_i})^2)^{\Delta_i}}.
\end{equation}
Any AdS $n$-star function is known to be obtained from a flat CFT $n$-star function by an analytic contiuation in the field dimensions $\{\Delta_i\}$ off the constraint of ``uniqueness''
\begin{equation} \label{4.11}
\sum_i \Delta_i = 4, \quad \text{with}\, 4 \,\text{being the spacetime dimension}
\end{equation}
and by a renormalization (see Appendix B). On the other hand, the flat CFT $n$-star functions for arbitrary spacetime dimension $d$ have been expressed by Mellin-Barnes integral representations that can be expanded in terms of hypergeometric functions of several variables \cite{symanzik}. An explicit expression for four-star functions with this method was given in \cite{lang2}. Using this result, we get for the special case
\begin{equation} \label{4.12}
\Delta, \Delta' \in \bf{N}, \quad \Delta'-\Delta \geq 0
\end{equation}
by factorizing a covariant multiplier
\begin{equation} \label{4.13}
D_{\Delta \Delta' \Delta \Delta'}(x_1,x_2,x_3,x_4) = (x_{12}^2)^{-\Delta} (x_{24}^2)^{\Delta-\Delta'} (x_{34}^2)^{-\Delta} G_{\Delta \Delta'}(u, v)
\end{equation}
where \footnote{The $\Psi$-function is defined as $\Psi(x)=\frac{d}{dx}ln \,\Gamma(x)$ (see 8.36 in \cite{grad}).}
\begin{align} \label{4.14}
G_{\Delta \Delta'}(u, v) &= \frac{\pi^2}{2} \frac{\Gamma(\Delta+\Delta'-2)}{\Gamma(\Delta)^2 \Gamma(\Delta')^2} \sum_{m=0}^\infty \frac{(1-v)^m}{m!}\{\sum_{n=0}^{\Delta'-\Delta-1} \frac{(-1)^n u^n}{n!} (\Delta'-\Delta-n-1)! \notag \\ &\times \frac{\Gamma(\Delta+n)^2 \Gamma(\Delta+n+m)^2}{\Gamma(2\Delta+2n+m)} +\sum_{n=\Delta'-\Delta}^\infty \frac{(-1)^{\Delta'-\Delta}u^n}{n!(n-\Delta'+\Delta)!} \frac{\Gamma(\Delta+n)^2\Gamma(\Delta+n+m)^2}{\Gamma(2\Delta+2n+m)} \notag \\
&\times[-log\, u + \Psi(n- \Delta'+\Delta+1) \notag \\ &+ \Psi(n+1) -2\Psi(\Delta+n) +2\Psi(2\Delta+2n+m)-2\Psi(\Delta+n+m)]\}.
\end{align}
  
In the analysis of the $\lambda = 8+2t$ towers we shall need analytic continuations of these functions off
\begin{equation} \label{4.15}
u \longrightarrow 0, \quad v \longrightarrow 1, \quad(\text{``Kummer relations''}).
\end{equation} 
First we turn to the graviton exchange graphs contributing to the $\lambda = 2$ tower. From \cite{dhoker2} we obtain
\begin{align} \label{4.16}
I_{grav}^{(t)} &= (\frac{6}{\pi^2})^4 (x_{12}^2x_{34}^2)^{-4} [16 (1+v-u)u^{-1} G_{45} +\frac{128}{9} (1+v)u^{-2} G_{35} \notag \\
&+\frac{32}{3} (1+v)u^{-3} G_{25} +18 G_{44} -\frac{46}{9} u^{-1} G_{34} -\frac{40}{9} u^{-2} G_{24} -\frac{8}{3} u^{-3} G_{14}].
\end{align}
Inserting the first finite power series from (\ref{4.14}) into (\ref{4.16}) we obtain the ``singular'' part
\begin{equation} \label{4.17}
I_{grav, sing}^{(t)} = 3 (\frac{6}{\pi^2})^3 (x_{12}^2x_{34}^2)^{-4}[u^{-3} S_3(v) + u^{-2} S_2(v) +u^{-1} S_1(v)]
\end{equation}
where
\begin{align}
S_3(v) &= \frac{8}{405}(1-v)^2 \, _2F_1(3,3;6;1-v), \label{4.18} \\
S_2(v) &= \frac{16}{27}\sum_{m=0}^\infty (m-4)(m+1)^2 \frac{(m+1)!}{(m+5)!} (1-v)^m, \label{4.19} \\
S_1(v) &= \frac{1}{105}(1+v) _2F_1(4,4;8;1-v) -\frac{2}{81}\, _2F_1(3,3;6;1-v).
\end{align}
For the analysis of these terms we use the results of section 2. In the case $\lambda = 2$ the $u$-factor in front of $\sum_{n,m}$ in (\ref{2.8}) is
\begin{equation} \label{4.21}
u^{\frac{1}{2}(\lambda-4-4)} = u^{-3} \quad \text{for}\, \lambda =2
\end{equation}
which fits to the $S_3$-terms in (\ref{4.17}). In fact, from (\ref{2.30})
\begin{equation} \label{4.22}
G_0(2,2;v) = \frac{4}{5!} (1-v)^2 \, _2F_1(3,3;6;1-v)
\end{equation}
so that,
\begin{equation} \label{4.23}
S_3(v) = \frac{16}{27} G_0(2,2;v).
\end{equation}
Consequently, we make the ansatz
\begin{equation} \label{4.24}
I_{grav, sing}^{(t)} = 3 (\frac{6}{\pi^2})^3 (x_{12}^2x_{34}^2)^{-4} u^{-3} \frac{16}{27} \sum_{n=0}^2 \frac{u^n}{n!} G_n(2,2;v).
\end{equation}
Given the values of $G_1(2,l;v)$ and $G_2(2,l;v)$ in the Appendix A, we can check the ansatz (\ref{4.24}) indeed.
Thus the only exceptional conformal block comes from the energy-momentum tensor, all other conserved tensor currents $(l \geq 4)$ have vanished in the strong coupling domain. On the other hand, in the free field limit $g_{YM} \rightarrow 0$ all these conserved tensor currents of even rank $l$ appear in the conformal partial wave decomposition with coupling constant $\gamma^{\frac{1}{2}}_l$ (\ref{3.20}). We can compare the coupling constants squared of the energy-momentum tensor to the image of the dilaton in the two limits
\begin{align}
\text{free field} &: \quad\frac{2}{9} \gamma_2 \frac{1}{N^2} = \frac{40}{3} \frac{1}{N^2}, \\
\text{strong coupling} &: \quad\frac{\pi^6}{72} 3 (\frac{6}{\pi^2})^3 \frac{16}{27} \frac{1}{N^2} = \frac{16}{3} \frac{1}{N^2}.
\end{align}
This difference is explained by the known splitting of the energy-momentum tensor at small coupling \cite{arutyunov}.
The conformal block of the energy-momentum tensor $(\lambda,l) = (2,2)$ can be subtracted from $I_{grav}^{(t)}$
\begin{equation} \label{4.27}
I_{grav}^{(t)}- 3 (\frac{6}{\pi^2})^3 (x_{12}^2x_{34}^2)^{-4} u^{-3} \frac{16}{27} \sum_{n=0}^\infty \frac{u^n}{n!} G_n(2,2;v).
\end{equation}
We shall see that the only conformal blocks $(\lambda,l)$ left in all Green functions are
\begin{equation} \label{4.28}
\lambda = 8 +2t, \quad t \in {\bf{N}}_0, \, l \in {\bf{N}}_0.
\end{equation}
These already appear at $\mathcal{O}(1)$ and obtain anomalous dimensions and corrected coupling constants at $\mathcal{O}(N^{-2})$. We shall now turn our attention to these ``composite fields''.

\section{AdS four-point functions and composite fields}
\setcounter{equation}{0}
The holographic images $\tilde{\Phi}$ of the dilaton and $\tilde{C}$ of the axion correspond to chiral primary operators of SYM$_4$
\begin{equation} \label{5.1}
\tilde{\Phi} \cong Tr (F^2), \quad \tilde{C} \cong Tr (F\tilde{F})
\end{equation}
where the trace is taken over the adjoint representation of the gauge group $SU(N)$. Conformal partial wave analysis of the AdS four-point functions leads besides the energy-momentum tensor only to composite fields of a pair of two fields $\tilde{\Phi}, \tilde{C}$. The corresponding conformal blocks are labelled by
\begin{equation} \label{5.2}
(\lambda,l, \Pi)
\end{equation}
where $\lambda$ is any
\begin{equation} \label{5.3}
\lambda = 8+2t, \quad t \in \bf{N}_0,
\end{equation}
$l$ is the tensor rank and $\Pi$ the parity, i.e. if
\begin{equation} \label{5.4}
\text{parity is} (-1)^l : \Pi = +1,
\quad\text{parity is} (-1)^{l+1} : \Pi = -1.
\end{equation}
From the bilocal fields
\begin{equation} \label{5.5}
\Psi_{\pm}(x_1, x_3) = \frac{1}{\sqrt{2}}[\pm \tilde{\Phi}(x_1)  \tilde{\Phi}(x_3)+ \tilde{C}(x_1)  \tilde{C}(x_3)]
\end{equation}
we find two towers of conformal blocks
\begin{equation} \label{5.6}
(\lambda,l,+1)_{\pm}, \quad l \, \text{even}
\end{equation}
and from
\begin{equation} 
\tilde{\Phi}(x_1)  \tilde{C}(x_3) \notag
\end{equation}
we obtain one tower
\begin{equation} \label{5.7}
(\lambda,l,-1), \quad l\, \text{even or odd}.
\end{equation}
The odd rank tensors are excluded from (\ref{5.6}) by Bose symmetry.
Our aim is to calculate the anomalous dimensions of these composite fields at order $\mathcal{O}(N^{-2})$. The fields themselves appear already at order $\mathcal{O}(1)$, namely by the decompositions of (\ref{4.5}), (\ref{4.7})
\begin{equation} \label{5.8}
1+v^{-4}=\sum_{l \in \bf{N}_0} \epsilon_l \frac{((l+3)!)^2}{(2l+7)!} (v-1)^l \, _2F_1(l+4, l+4; 2l+8;1-v)
\end{equation}
with
\begin{equation} \label{5.9}
\epsilon_l =
\begin{cases}
\frac{1}{18} (2l+7)(l+1)_6, \quad l \; \text{even}  \\
0, \quad l \; \text{odd}
\end{cases}
\end{equation}
and
\begin{equation} \label{5.10}
1 = \sum_{l \in \bf{N}_0} \tilde{\epsilon}_l \frac{((l+3)!)^2}{(2l+7)!} (v-1)^l \, _2F_1(l+4, l+4; 2l+8;1-v)
\end{equation}
with
\begin{equation} \label{5.11}
\tilde{\epsilon}_l = \frac{1}{36} (2l+7)(l+1)_6.
\end{equation}
Note that $\tilde{\epsilon}_l$ is the square of a coupling constant.

Next we extract the coefficient functions of $-log\, u$ at order $\mathcal{O}(N^{-2})$. For any of the integrals involved, we denote them
\begin{equation} \label{5.12}
I = 3 (\frac{6}{\pi^2})^3 [a(v) (-log\, u) +b(v)]+ \mathcal{O}(u).
\end{equation}
This gives information only on the $\lambda=8$ towers. Then we get: ($3 (\frac{6}{\pi^2})^3 \frac{\pi^6}{72} = 9$)
\begin{align}
A_1=D_1&: \quad 9[a_{grav}^{(s)}(v)+a_{grav}^{(t)}(v)+a_{grav}^{(u)}(v)], \label{5.13}\\
B_1=C_1&: \quad 9[a_{grav}^{(t)}(v)+a^{(s+u+q)}(v)], \label{5.14}\\
E_1 &: \quad9 a^{(e)}(v) \label{5.15}
\end{align}
where $a_{grav}^{(e)}(v)$ is calculated from the $C$-exchange graph in the $t$-channel whose $s$- and $u$-channel versions appear in Fig.4.

Fields have well-defined anomalous dimensions if they are orthogonal w.r.t. the two-point functions. The matrix
\begin{equation} \label{5.16}
\begin{pmatrix} A & B \\ C & D \end{pmatrix} = <\binom{\tilde{\Phi}(x_1) \tilde{\Phi}(x_3)}{\tilde{C}(x_1)\tilde{C}(x_3)} (\tilde{\Phi}(x_2) \tilde{\Phi}(x_4), \tilde{C}(x_2) \tilde{C}(x_4))>
\end{equation}
is proportional to the unit matrix (see (\ref{4.5}), (\ref{4.6})) at order $\mathcal{O}(1)$ but this degeneracy is lifted at order $\mathcal{O}(N^{-2})$
\begin{equation} \label{5.17}
\begin{pmatrix}A & B \\ C & D\end{pmatrix} = \begin{pmatrix}A_0 & 0 \\ 0 & A_0 \end{pmatrix} + \frac{1}{N^2} \begin{pmatrix}A_1 & B_1 \\ B_1 & A_1 \end{pmatrix} + \mathcal{O}(N^{-4}).
\end{equation}
Diagonalization (at order $\mathcal{O}(N^{-2})$) can be achieved by the matrix 
\begin{equation} \label{5.18}
u = \frac{1}{\sqrt{2}} \begin{pmatrix}1 & -1\\ 1 & 1\end{pmatrix}
\end{equation}
so that
\begin{equation} \label{5.19}
u^{T}\begin{pmatrix}A & B \\ C & D\end{pmatrix} u = \begin{pmatrix}A_0 & 0 \\ 0 & A_0 \end{pmatrix} + \frac{1}{N^2}\begin{pmatrix} A_1+B_1 & 0 \\ 0 & A_1-B_1\end{pmatrix} +  \mathcal{O}(N^{-4})
\end{equation}
and
\begin{equation} \label{5.20}
( \tilde{\Phi}(x_2) \tilde{\Phi}(x_4), \tilde{C}(x_2)\tilde{C}(x_4)) u = (\Psi_+(x_2, x_4), \Psi_-(x_2, x_4)).
\end{equation}

For the partial wave decomposition we expand $A_1 \pm B_1$ and $E_1$ (see (\ref{5.13})-(\ref{5.15}))
\begin{align}
&9[a_{grav}^{(s)}(v)+2 a_{grav}^{(t)}(v)+ a_{grav}^{(u)}(v)+ a^{(s+u+q)}(v)]=\sum_{l \in \bf{N}_0} \zeta_{l,+} G_0(8,l;v), \label{5.21}\\
&9[a_{grav}^{(s)}(v)+ a_{grav}^{(u)}(v)-  a^{(s+u+q)}(v)]=\sum_{l \in \bf{N}_0} \zeta_{l,-} G_0(8,l;v), \label{5.22}\\
&9 a^{(e)}(v) = \sum_{l \in \bf{N}_0} \vartheta_l G_0(8,l;v). \label{5.23}
\end{align}
Inserting the correct exponent into the $u$-power of (\ref{2.8}), we get
\begin{equation} \label{5.24}
u^{\frac{1}{2}(8+l+\eta(8,l)-8-l)} = 1 + \frac{1}{2} \eta(8,l) log \,u + ...
\end{equation}
so that the anomalous dimensions are
\begin{align} \label{5.25}
\eta(8,l,+)_\pm &= -2 \frac{\zeta_{l,\pm}}{\epsilon_l} \frac{1}{N^2},\notag \\
\eta(8,l,-) &= -2 \frac{\vartheta_{l}}{\tilde{\epsilon}_l} \frac{1}{N^2}.
\end{align}

The main issue left is to derive the functions $a(v)$ using analytic continuation of generalized hypergeometic functions. 
From (\ref{4.14}), (\ref{4.16}) we obtain
\begin{equation} \label{5.26}
 a_{grav}^{(t)}(v) = \frac{-251}{3^5 \cdot 35} (1+v) \, _2F_1(5,5;10; 1-v) + \frac{1277}{3^4 \cdot 35} \, _2F_1(4,4;8; 1-v).
\end{equation}
We insert (\ref{B.18}), (\ref{B.19}) into (\ref{4.16}) and get
\begin{align} \label{5.27}
I^{(s)}_{grav} =& (\frac{6}{\pi^2})^4 (x_{12}^2 x_{34}^2)^{-4} [16(u+v-1) u^{-4} G_{45}(\frac{1}{u},\frac{v}{u}) \notag \\
&+\frac{128}{9} (u+v) u^{-3} G_{35}(\frac{1}{u},\frac{v}{u}) + \frac{32}{3} (u+v) u^{-2} G_{25}(\frac{1}{u},\frac{v}{u}) \notag \\
&+ 18 G_{44}(\frac{1}{u},\frac{v}{u})- \frac{46}{9} u^{-3} G_{34}(\frac{1}{u},\frac{v}{u}) \notag \\
&- \frac{40}{9} u^{-2} G_{24}(\frac{1}{u},\frac{v}{u}) - \frac{8}{3} u^{-1} G_{14}(\frac{1}{u},\frac{v}{u})]. 
\end{align}
Making use of (\ref{B.25}) yields
\begin{align} \label{5.28}
 a_{grav}^{(s)}(v)=& -\frac{2}{7} (1-v) \, _2F_1(4,5;9;1-v) + \frac{64}{189} v \, _2F_1(3,5;8;1-v) \notag \\
&+\frac{16}{45} v \, _2F_1(2,5;7;1-v) + \frac{3}{7} \,  _2F_1(4,4;8;1-v) \notag \\
&-\frac{23}{135} \, _2F_1(3,4;7;1-v) - \frac{2}{9} ( _2F_1(2,4;6;1-v) + \,  _2F_1(1,4;5;1-v)).
\end{align}
In the same fashion we get from (\ref{B.21}), (\ref{B.22}) and (\ref{4.16})
\begin{align} \label{5.29}
I^{(u)}_{grav} =& (\frac{6}{\pi^2})^4 (x_{12}^2 x_{34}^2)^{-4} [16(u-v+1) v^{-1} G_{45}(v, u) \notag \\
&+\frac{128}{9} (1+u) v^{-2} G_{35}(v, u) + \frac{32}{3} (1+u) v^{-3} G_{25}(v, u) \notag \\
&+ 18 G_{44}(v, u)- \frac{46}{9} v^{-1} G_{34}(v, u)) \notag \\
&- \frac{40}{9} v^{-2} G_{24}(v, u) - \frac{8}{3} v^{-3} G_{14}(v, u)] 
\end{align}
which by (\ref{B.26}) leads to
\begin{align} \label{5.30}
 a_{grav}^{(u)}(v)=& \frac{2}{7} (1-v) \, _2F_1(5,5;9;1-v) + \frac{64}{189}  \, _2F_1(5,5;8;1-v) \notag \\
&+\frac{16}{45}  \, _2F_1(5,5;7;1-v) + \frac{3}{7} \, _2F_1(4,4;8;1-v) \notag \\
&-\frac{23}{135} \, _2F_1(4,4;7;1-v) - \frac{2}{9} ( _2F_1(4,4;6;1-v) + \, _2F_1(4,4;5;1-v)).
\end{align}
In the case of $B_1=C_1$ we get a contribution of the graphs $s, u, q$ in Fig.4 which have been evaluated
 and give 
\begin{equation} \label{5.31}
I^{(s+u+q)}=  (\frac{6}{\pi^2})^4 (x_{12}^2 x_{34}^2)^{-4} [64 G_{45}(u, v) -32 G_{44}(u, v)]
\end{equation}
from which we obtain with (\ref{4.14})
\begin{equation} \label{5.32}
a^{(s+u+q)}(v)= -\frac{16}{21} \, _2F_1(4,4;8;1-v).
\end{equation}
By application of the analytic continuation
\begin{equation} 
x_2 \longleftrightarrow x_3 \notag
\end{equation}
we derive $E_1$ from $B_1$ so that
\begin{equation} \label{5.33}
E_1: \quad I^{(s)}_{grav} +  (\frac{6}{\pi^2})^4 (x_{12}^2 x_{34}^2)^{-4} [64 u^{-4} G_{45}(\frac{1}{u}, \frac{v}{u}) -32 u^{-4} G_{44}(\frac{1}{u},\frac{v}{u})]
\end{equation}
or 
\begin{equation} \label{5.34}
a^{(e)}(v)=  a_{grav}^{(s)}(v)+\frac{8}{7} \, _2F_1(4,5;9;1-v)-\frac{16}{21} \, _2F_1(4,4;8;1-v).
\end{equation}

We will try to solve (\ref{5.21})-(\ref{5.23}) systematically for all $l$. For this purpose we expand
\begin{align} 
 a_{grav}^{(s)}(v)&= \sum_{m=0}^\infty A_m^{(s)} \frac{(1-v)^m}{m!}, \label{5.35} \\
 a_{grav}^{(u)}(v)&= \sum_{m=0}^\infty A_m^{(u)} \frac{(1-v)^m}{m!}. \label{5.36}
\end{align}
We find from (\ref{5.28}), (\ref{5.30})
\begin{align} 
A_m^{(s)} &= -\frac{16}{63} \frac{m-2}{m+1} \frac{((4)_m)^2}{(8)_m}, \label{5.37} \\
A_m^{(u)} &= \frac{2}{945} (m+2)(m^2+19m+120) \frac{((4)_m)^2}{(8)_m}. \label{5.38}
\end{align}
Then by expanding $G_0(8,l;v)$ and reordering factors, we get for (\ref{5.22}), (\ref{5.23})
\begin{align}
&\sum_{l \in \bf{N}_0} \zeta_{l,-} \frac{(-1)^l m! (m+7)!}{(m-l)! (m+l+7)!} = 320 \frac{2m+5}{m+1} + \frac{8}{3} (m+2)(m^2+19m+120), \label{5.39} \\
&\sum_{l \in \bf{N}_0} \vartheta_l \frac{(-1)^l m! (m+7)!}{(m-l)! (m+l+7)!} = 320 \frac{5m^2+12m+28}{(m+1)(m+8)}. \label{5.40}
\end{align}
Only the ratios
\begin{equation} \label{5.41}
\frac{\zeta_{l,-}}{\epsilon_l}, \quad \frac{\vartheta_l}{\tilde{\epsilon}_l}
\end{equation}
enter the anomalous dimensions (\ref{5.25}). From (\ref{5.40}) we obtain
\begin{equation} \label{5.42}
\vartheta_l = \frac{4}{3} (2l+7)(l+2)_4 - 1440 \, \delta_{l,1}
\end{equation}
and 
\begin{equation} \label{5.43}
\frac{\vartheta_l}{\tilde{\epsilon}_l} = \frac{48}{(l+1)(l+6)} -\frac{8}{7} \, \delta_{l,1}.
\end{equation}
The proof that (\ref{5.42}) solves (\ref{5.40}) is presented in Appendix C.
For the other ratio we get
\begin{equation} \label{5.44}
\frac{\zeta_{l,-}}{\epsilon_l} =
\begin{cases}
 \frac{\vartheta_l}{\tilde{\epsilon}_l}, \quad l \, \text{even} \\
0, \quad  l \, \text{odd}
\end{cases}
\end{equation}
which was quoted in the Introduction as eqn.(1.8). The normalization of $\Psi_0$ is
\begin{equation} \label{5.45}
\Psi_0(x_1,x_3)=\frac{1}{\sqrt{2}}[\tilde{\Phi}(x_1)\tilde{C}(x_3)+\tilde{C}(x_1)\tilde{\Phi}(x_3)].
\end{equation}

Finally we mention that for 
\begin{equation}
 a_{grav}^{(t)}(v) \notag
\end{equation}
we have only calculated
\begin{equation} \label{5.46}
 a_{grav}^{(t)}(1) = A_0^{(t)} = \frac{3329}{3^5\cdot 35}
\end{equation}
which yields the anomalous dimension
\begin{equation} \label{5.47}
\eta(8,0,+)_+ = -2 \frac{4409}{27 \cdot 35} \frac{1}{N^2}.
\end{equation}

\begin{appendix}
\section{ The coefficients of the functions $G_n(2,l;v)$}
\setcounter{equation}{0}
We start from the definition (\ref{2.29}) which for $\lambda=2, d=4$ obtains a contribution only for $M=l$. We insert (\ref{2.32}) and perform the $t$-summation. Since a general formula for $G_0$ exists in (\ref{2.30}), we need $G_n$ only for $n \geq 1$. We obtain
\begin{align} \label{A.1}
n=1: 
\quad \tilde{A}_{1m}^{(l)} &= \frac{(-1)^l}{2} \{\frac{m! ((m+1)!)^2}{(m+1-l)!(m+2+l)!} [(m+1)-1] \notag \\
&+\frac{m!(m+1)!(m+2)!}{(m+2-l)!(m+3+l)!}[-(m+1)_2+ 2(m+1)-(l-1)(l+2)]\},
\end{align}
\begin{align} \label{A.2}
n=2: \quad
\tilde{A}_{2m}^{(l)} &= \frac{(-1)^l}{2}\{\frac{m! ((m+2)!)^2}{(m+2-l)!(m+3+l)!} [(m+1)_2+2] \notag \\
&+\frac{m!(m+2)!(m+3)!}{(m+3-l)!(m+4+l)!}[-2(m+1)_3 -12(m+1)+4(l-2)(l+3)] \notag \\
&+\frac{m!(m+2)!(m+4)!}{(m+4-l)!(m+5+l)!} \notag \\ &\times[(m+1)_4 +12(m+1)_2-8(m+1)(l-2)(l+3)+(l-3)_2(l+3)_2]\},
\end{align}
\begin{align} \label{A.3}
n=3: \quad
\tilde{A}_{3m}^{(l)} &= \frac{(-1)^l}{2} \{\frac{m! ((m+3)!)^2}{(m+3-l)!(m+4+l)!} [(m+1)_3 -6] \notag \\
&+\frac{m!(m+3)!(m+4)!}{(m+4-l)!(m+5+l)!}[-3(m+1)_4 +72(m+1)-18(l-3)(l+4)] \notag \\
&+\frac{m!(m+3)!(m+5)!}{(m+5-l)!(m+6+l)!}[3(m+1)_5 -180(m+1)_2 \notag \\ &+90(m+1)(l-3)(l+4)-9(l-4)_2(l+4)_2] \notag \\
&+\frac{m!(m+3)!(m+6)!}{(m+6-l)!(m+7+l)!}[-(m+1)_6 +120(m+1)_3 \notag \\ &-90(m+1)_2(l-3)(l+4)+18(m+1)(l-4)_2(l+4)_2-(l-5)_3(l+4)_3]\}. 
\end{align}

\section{Analytic continuation of generalized hypergeometric functions}
\setcounter{equation}{0}
A vertex of $m$ scalar fields in AdS$_{D+1}$ conformal field theory defines a Green function ($x_i \in {\bf{R}}_D = \partial\text{AdS}_{D+1}$)
\begin{equation} \label{B.1}
\Gamma(x_1,...,x_m)= \pi^{-\frac{D}{2}} \int_{z_0>0} \frac{d^{D+1}}{z_0^{D+1}} \prod_{i=1}^m (\frac{z_0}{z_0^2 + (\vec{z}-\vec{x_i})^2})^{\alpha_i}
\end{equation}
where$\{\alpha_i\}$ are field dimensions constrained only by 
\begin{equation} \label{B.2}
\alpha_i \geq \frac{1}{2} D -1.
\end{equation} 
In $\bf{R}_D$ conformal field theory such a vertex is given by
\begin{equation} \label{B.3}
G_m(x_1,...,x_m) = \pi^{-\frac{D}{2}} \int_{\bf{R}_D} d^Dy \prod_{i=i}^m ((y-x_i)^2)^{-\alpha_i}
\end{equation}
where conformal covariance necessitates ``uniqueness''
\begin{equation} \label{B.4}
\sum_{i=1}^m \alpha_i = D.
\end{equation}
Nevertheless $\Gamma_m$ can be obtained from $G_m$ by an analytic continuation and a renormalization. Since all $G_m$ are known to be expressible as Mellin-Barnes integrals which are easily expanded as hypergeometric functions this fact is of quite an importance.
To prove this assertion, we express $G_m$ by an integral
\begin{equation} \label{B.5}
G_m(D;\alpha_1,...,\alpha_m) = [\prod_{i=1}^m \Gamma(\alpha_i)]^{-1}(\prod_{i=1}^m \int_0^\infty dt_i \, t_i^{\alpha_i-1}) \, T^{-\frac{D}{2}} e^{-\frac{1}{T} \sum_{i<j}t_it_j x_{ij}^2}
\end{equation}
with
\begin{equation} \label{B.6}
T = \sum_{i=1}^m t_i.
\end{equation}
Though $G_m$ is defined only on the manifold (\ref{B.4}), we can use the representation (\ref{B.5}) to continue it off this manifold so that $D$ and $\{\alpha_i\}$ are independent variables. With the same auxiliary integration as those leading to (\ref{B.5}) and integrating $z_0$, we get
\begin{equation} \label{B.7}
\Gamma_m(D;\alpha_1,...,\alpha_m) = \frac{1}{2} \Gamma(\frac{1}{2}(\sum_{i=1}^m \alpha_i - D)) \, G_m(\sum_{i=1}^m \alpha_i;\alpha_1,...,\alpha_m).
\end{equation}
The $D$-dependence of $\Gamma_m$ stems solely from the normalizing $\Gamma$-function.

From now on we consider the function $G_m$ as being defined by (\ref{B.5}), (\ref{B.6}) with unconstrained field dimensions but with the substitution (\ref{B.4}) as indicated in (\ref{B.7}). We study the case $m=4$. With Symanzik's technique of evaluating (\ref{B.5}) and expanding the Mellin-Barnes integral, we get
\begin{equation} \label{B.8}
G_m(x_1, x_2, x_3, x_4) = [\prod_{i=1}^m\Gamma(\alpha_i)]^{-1} r_{12}^{\alpha_4-\frac{1}{2}D} r_{14}^{\frac{1}{2}D-\alpha_1-\alpha_4} r_{24}^{\frac{1}{2}D-\alpha_2-\alpha_4} r_{34}^{-\alpha_3} F(u,v)
\end{equation}
where
\begin{equation} \label{B.9}
r_{ij} = (x_{ij})^2, \quad u = \frac{r_{13} r_{24}}{r_{12} r_{34}}, \quad  v = \frac{r_{14} r_{23}}{r_{12} r_{34}}
\end{equation}
and
\begin{align} \label{B.10}
F(u,v) = & u^{-\frac{1}{2}(\alpha_1 + \alpha_3)} \sum_{n,m=0}^\infty \frac{u^n (1-v)^m}{n! m!} \{u^{\frac{1}{2}(\alpha_1 + \alpha_3)} \Gamma(\frac{1}{2}(\alpha_2+\alpha_4-\alpha_1-\alpha_3)) \notag \\
&\times \frac{\Gamma(\alpha_1+n) \Gamma(\frac{1}{2}D-\alpha_2+n) \Gamma(\alpha_3+n+m) \Gamma(\frac{1}{2}D-\alpha_4+n+m)}{\Gamma(\alpha_1+\alpha_3+2n+m)(\frac{1}{2}(\alpha_1+\alpha_3-\alpha_2-\alpha_4)+1)_n}+ u^{\frac{1}{2}(\alpha_2 + \alpha_4)} \notag \\ & \times\Gamma(\frac{1}{2}(\alpha_1+\alpha_3-\alpha_2-\alpha_4)) \frac{\Gamma(\alpha_4+n) \Gamma(\frac{1}{2}D-\alpha_3+n) \Gamma(\alpha_2+n+m) \Gamma(\frac{1}{2}D-\alpha_1+n+m)}{\Gamma(\alpha_2+\alpha_4+2n+m)(\frac{1}{2}(\alpha_2+\alpha_4-\alpha_1-\alpha_3)+1)_n}.
\end{align}
This double series converges in a neighborhood of 
\begin{equation} \label{B.11}
u = 0, \quad v=1
\end{equation}
and is suited for an operator product expansion in the ``$t$-channel''
\begin{equation} \label{B.12}
r_{13} \longrightarrow 0, \quad r_{24} \longrightarrow 0.
\end{equation}
Correspondingly, the  ``$u$-channel'' is 
\begin{equation} \label{B.13}
r_{14} \longrightarrow 0, \quad r_{23} \longrightarrow 0
\end{equation}
and the ``$s$-channel''
\begin{equation} \label{B.14}
r_{12} \longrightarrow 0, \quad r_{34} \longrightarrow 0.
\end{equation}
The vertex function $G_m$ is invariant under permutations of the set of pairs
\begin{equation} 
\{(x_1, \alpha_1),(x_2, \alpha_2), (x_3, \alpha_3),..., (x_m, \alpha_m)\} \notag
\end{equation}
but the integration technique has hidden this symmetry. For $m=4$ the symmetry group $S_4$ has a subgroup
\begin{equation}
Z_2^{(1)} \times Z_2^{(2)} \notag
\end{equation}
which leaves $u$ and $v$ invariant. In particular
\begin{equation} \label{B.15}
Z_2^{(1)}=\{(1,2,3,4), (4,3,2,1)\}
\end{equation}
acts on the hypergeometric series (\ref{B.8}), (\ref{B.10}) trivially. The other subgroup
\begin{equation} \label{B.16}
Z_2^{(2)}=\{(1,2,3,4), (3,4,1,2)\}
\end{equation}
produces Euler identities for the Gaussian hypergeometric functions involved in (\ref{B.10}), e.g.
\begin{multline} \label{B.17}
_2F_1(\alpha_3+n, \frac{1}{2}D-\alpha_4+n; \alpha_1+\alpha_3+2n; 1-v)\\ = v^{\alpha_1+\alpha_4-\frac{1}{2}D}\, _2F_1(\alpha_1+n, \frac{1}{2}D-\alpha_2+n; \alpha_1+\alpha_3+2n; 1-v)
\end{multline}
(see \cite{grad}, eqn.9.131.1).

The other classes of $S_4/(Z_2 \times Z_2)$ produce analytic continuation formulae. They can also be directly derived from the Kummer relations for $_2F_1$ (e.g. in \cite{grad} eqns. 9.132.1, 9.132.2, but this set of equations is not complete). These continuations give formulae useful for operator product expansions in the $s$-channel and $u$-channel.
Define $g_{t \rightarrow s}: (1,2,3,4) \rightarrow (1,3,2,4)$. Then this element acts on $u$ and $v$ as
\begin{align} 
u \underset{g_{t \rightarrow s}}{\longrightarrow} u'&= \frac{1}{u},  \label{B.18} \\
v \underset{g_{t \rightarrow s}}{\longrightarrow} v'&= \frac{v}{u}.  \label{B.19}
\end{align}
If we denote the function
\begin{equation} 
F'(u,v) \notag
\end{equation}
as the result of the replacement of $\alpha_2$ be $\alpha_3$ in $F(u,v)$, we get as identity
\begin{equation} \label{B.20}
F'(u',v') = u^{\frac{1}{2}D-\alpha_4} F(u,v).
\end{equation}
Define $g_{t \rightarrow u}: (1,2,3,4) \rightarrow (1,2,4,3)$. Then this element acts on $u$ and $v$ as
\begin{align}
u \underset{g_{t \rightarrow u}}{\longrightarrow} u''&= v,  \label{B.21} \\
v \underset{g_{t \rightarrow u}}{\longrightarrow} v''&= u.  \label{B.22}
\end{align}
Consequently if we denote the substitution
\begin{equation}
F \underset{\alpha_3 \leftrightarrow \alpha_4}{\longrightarrow} F'' \notag
\end{equation}
we obtain
\begin{equation} \label{B.23}
F''(u'',v'') = u^{\frac{1}{2}(\alpha_1+\alpha_3-\alpha_2-\alpha_4)}  v^{\frac{1}{2}(\alpha_2+\alpha_3-\alpha_1-\alpha_4)}  F(u,v).
\end{equation}
Now we specialize to the case (\ref{4.12}), (\ref{4.13}).
In the case $G_{\Delta \Delta'}(u,v)$ we cannot simply insert (\ref{4.12}), (\ref{4.13}) since this leads to undefined expressions. But a deformation
\begin{equation} \label{B.24}
\Delta' = \Delta +k+\epsilon, \quad k \in \bf{N}_0
\end{equation}
enables us to cancel the poles in $\epsilon$ before we let $\epsilon$ go to zero. By this technique, we get (\ref{4.14}).

Next we need analytic continuations of $G_{\Delta \Delta'}(u,v)$ to the $s$- and $u$-channels. We express $G_{\Delta \Delta'}(u',v')$ and  $G_{\Delta \Delta'}(u'',v'')$ (see (\ref{B.18}),(\ref{B.19}) and  (\ref{B.21}), (\ref{B.22})) as power series in $u$ and $1-v$, namely
\begin{align} \label{B.25}
G_{\Delta \Delta'}(u', v') =& \frac{\pi^2}{2} \frac{\Gamma(\Delta+\Delta'-2)}{\Gamma(\Delta)^2 \Gamma(\Delta')^2} u^{\Delta} \sum_{n,m=0}^\infty \frac{u^n (1-v)^m}{(n!)^2 m!} \notag \\
&\times \frac{\Gamma(\Delta+n) \Gamma(\Delta'+n)\Gamma(\Delta+n+m)\Gamma(\Delta'+n+m)}{\Gamma(\Delta+\Delta'+2n+m)} \{-log \, u +2 \Psi(n+1)-\Psi(\Delta+n)\notag \\&-\Psi(\Delta'+n)-\Psi(\Delta+n+m)-\Psi(\Delta'+n+m)+2\Psi(\Delta+\Delta'+2n+m)\}
\end{align}
and 
\begin{align} \label{B.26}
G_{\Delta \Delta'}(u'', v'') =& \frac{\pi^2}{2} \frac{\Gamma(\Delta+\Delta'-2)}{\Gamma(\Delta)^2 \Gamma(\Delta')^2} v^{\Delta'-\Delta} \sum_{n,m=0}^\infty \frac{u^n (1-v)^m}{n! m!} \notag \\
&\times \frac{\Gamma(\Delta+n)^2 \Gamma(\Delta'+n+m)^2}{\Gamma(\Delta+\Delta'+2n+m)} \{-log \, u +2 \Psi(n+1)-2\Psi(\Delta+n)-2\Psi(\Delta'+n+m) \notag \\
&+2\Psi(\Delta+\Delta'+2n+m)\}.
\end{align}
Instead of exploiting the symmetry of the star graph we were also able to derive these equations by repeated application of the Kummer relations (three term relations) for the Gaussian hypergeometric function $_2F_1$. We propose to denote all relations between multi-variable hypergeometric functions that can be derived from the symmetry of graphs as ``generalized Kummer relations''.

\section{Proof of equation (5.42)}
\setcounter{equation}{0}
We decompose $\vartheta_l$ into
\begin{equation} \label{C.1}
\vartheta_l = \vartheta_l' +\vartheta_l'',\quad \vartheta_l'= -1440 \,\delta_{l,1}
\end{equation}
and insert $\vartheta_l'$ into (\ref{5.40}). We obtain
\begin{equation} \label{C.2}
\sum_{l \in \bf{N}_0} \vartheta_l'' \frac{(-1)^l m! (m+7)!}{(m-l)!(m+l+7)!}\overset{?}{=} 160 \frac{(m+7)}{(m+1)}.
\end{equation}
Then we insert $\vartheta_l''$ in the l.h.s.
\begin{equation} \label{C.3}
\sum_{l=0}^\infty (2l+7)(l+2)_4 \frac{(-m)_l}{(m+8)_l} \overset{?}{=} 120 \frac{(m+7)}{(m+1)}.
\end{equation}
We want to make use of 
\begin{align} \label{C.4}
\sum_{l=0}^\infty \frac{(l+1)_k (-m)_l}{(m+8)_l} &= k!\, _2F_1(k+1,-m;m+8;1) \notag \\
&= k! \frac{(m+7)!(2m-k+6)!}{(2m+7)!(m-k+6)!}.
\end{align}
Therefore we expand
\begin{equation} \label{C.5}
(2l+7)(l+2)_4 = \sum_{k=0}^5 a_k (l+1)_k
\end{equation}
with
\begin{equation} \label{C.6}
a_k = \frac{5!}{k!}(1+\delta_{k,5})
\end{equation}
and have to prove finally
\begin{equation} \label{C.7}
\sum_{k=0}^5 (1+\delta_{k,5}) \frac{(m+7)!(2m-k+6)!}{(2m+7)!(m-k+6)!} \overset{?}{=} \frac{(m+7)}{(m+1)}.
\end{equation}
The sum is decomposed into the three terms:
\begin{align} 
&(1) \quad \sum_{k=0}^\infty \frac{(m+7)!(2m-k+6)!}{(2m+7)!(m-k+6)!} =  \frac{(m+7)}{(m+1)}, \\
&(2) \quad \frac{(m+7)!(2m-k+6)!}{(2m+7)!(m-k+6)!} \arrowvert_{k=5} = \frac{(m+7)!(2m+1)!}{(2m+7)!(m+1)!},\\
&(3) \quad - \sum_{k=6}^\infty \frac{(m+7)!(2m-k+6)!}{(2m+7)!(m-k+6)!} = - \frac{(m+7)!(2m+1)!}{(2m+7)!(m+1)!}.
\end{align}
The series $(1)$ and $(3)$ are Gaussian hypergeometric. Cancelling $(2)$ and $(3)$, the desired result arises from $(1)$.

\end{appendix}

\end{document}